\documentstyle[twoside,fleqn,espcrc2]{article}

\newcommand{\AmS}{{\protect\the\textfont2
  A\kern-.1667em\lower.5ex\hbox{M}\kern-.125emS}}

\title{Three-Family Grand Unification in String Theory
           \thanks{Report-no: HUTP-97/A042, NUB 3166.}
           \thanks{Talk presented at SUSY'97.}}

\author{Zurab Kakushadze
      \address{Lyman Laboratory of Physics, Harvard University, Cambridge, MA, 02138} 
      \address{Department of Physics, Northeastern University, Boston, MA 02115}
        \thanks{E-mail: zurab@string.harvard.edu.}}

\begin{document}

\begin{abstract}
We briefly discuss the status of three-family grand unified string models. 
\end{abstract}

\maketitle

{}The outstanding question of string theory is whether it describes {\em our} universe. This question turns out to be very difficult to answer as the space of classical string vacua has a very large degeneracy, and there lack objective criteria that would select a particular string vacuum among the numerous possibilities. One might expect non-perturbative string dynamics to lift, partially or completely, this degeneracy in the moduli space. {\em A priori} it is reasonable to suspect that non-perturbative dynamics may not select the unique vacuum, but rather pick out 
a number of consistent vacua with completely broken supersymmetry. The reader might find the following analogy, that belongs to Henry Tye, amusing. The ancient Greeks, among many others, believed that the earth was the center of the universe, and everything else revolved around it. Just as this perception did not prove correct, it might be too naive to believe that the only truly consistent superstring vacuum is the one that describes our universe. If so, then to find the string vacuum where we live, we would need to impose some additional, namely, phenomenological constraints. This approach has been known as ``superstring phenomenology''. The latter must still be augmented with understanding of non-perturbative dynamics as  superstring is believed not to break supersymmetry perturbatively.   

{}The guiding principle of superstring phenomenology is very simple: it is to find a string model that within its effective low energy field theory will reproduce the Standard Model of strong and electroweak interactions. It is, however, not known how to fully embed the latter into string theory with all of its complexity, so one is bound to try to incorporate only a few phenomenologically desirable features at a time (such as, say, the gauge group, number of families, {\em etc.}). This ultimately leads to numerous possibilities for embedding the Standard Model in superstring that {\em a priori} seem reasonable. To make progress along these lines one, therefore, needs as restrictive constraints as possible.

{} The main arena for model building within the context of superstring phenomenology has been 
perturbative heterotic superstring. The reason for this is that such model building is greatly facilitated by existence of relatively simple rules (such as free-fermionic \cite{FFC} and orbifold \cite{Orb,NonAbe} constructions). Moreover, many calculational tools (such as, say, scattering amplitudes and rules for computing superpotentials \cite{scatt})
are either readily available, or can be developed for certain cases of interest. Despite enormous progress made in the past few years in understanding non-perturbative superstring vacua, the state of the art there is still far from being competitive with perturbative heterotic superstring, and tools available in the latter framework  must be first generalized to include the former before superstring phenomenology can step into this new terrain.

{}Thus, to be specific, let us concentrate on perturbative heterotic superstring. Within this framework the total rank of the gauge group (for $N=1$ space-time supersymmetric models) is 22 or less. After accommodating the Standard Model of strong and electroweak interactions (with gauge group $SU(3)_c \otimes SU(2)_w \otimes U(1)_Y$ whose rank is 4), the left-over rank for the hidden and/or horizontal gauge symmetry is 18 or less. The possible choices here are myriad \cite{standard} and largely unexplored. The situation is similar for embedding unification within a {\em semi-simple} \cite{semi} gauge group $G\supset SU(3)_c \otimes SU(2)_w \otimes U(1)_Y$ ({\em e.g.}, $SU(5)\otimes U(1)$).

{}The state of affairs is quite different if one tries to embed grand unification within a {\em simple} gauge group  $G\supset SU(3)_c \otimes SU(2)_w \otimes U(1)_Y$. Thus, in effective field theory to break the grand unified gauge group $G$ down to that of the Standard Model an adjoint 
(or some other appropriate higher dimensional) Higgs field must be present among the light degrees of freedom. In perturbative heterotic superstring such states in the massless spectrum are compatible with $N=1$ supersymmetry and chiral fermions only if the grand unified gauge group is realized via a current algebra at level $k>1$ \cite{Lew}. This ultimately leads to reduction of total rank of the gauge group, and, therefore, to smaller room for hidden/horizontal symmetry, which greatly limits the number of possible models.  

{}The simplest way to obtain higher-level models is via the following construction. Start from a $k$-fold product $G\otimes G\otimes \cdots \otimes G$ of the grand unified gauge group $G$ realized via a level-1 current algebra. The diagonal subgroup $G_{diag}\subset  G\otimes G\otimes \cdots \otimes G$ then is realized via level $k$ current algebra. (Note that in carrying out this procedure the rank of the gauge group is reduced from $kr$ to $r$, the latter being the rank of $G$.) As far as the Hilbert space is concerned, here we are identifying the states under the ${\bf Z}_k$ cyclic symmetry of the $k$-fold product $G\otimes G\otimes \cdots \otimes G$. This is nothing but ${\bf Z}_k$ orbifold action, namely, modding out by the outer automorphism.

{}An immediate implication of the above construction is a rather limited number of possibilities. Thus, for a grand unified gauge group $G=SO(10)$ with, say, $k=3$, the left-over rank (for the hidden and/or horizontal gauge symmetry) is at most 7 ($=22-3\times 5$). This is to be compared with the left-over rank 18 in the case of the Standard Model embedding. Taking into account that the number of models grows (roughly) as a factorial of the left-over rank, it becomes clear that grand unified model building is much more restricted than other embeddings.

{}Since desired adjoint (or higher dimensional) Higgs fields are allowed already at level $k=2$, multiple attempts have been made in the past several year to construct level-2 grand unified string models \cite{two}. None of them, however, have yielded 3-family models. There is no formal proof that 3-family models cannot be obtained from level-2 constructions, but one can intuitively understand why attempts to find such models have failed. In the $k=2$ construction the orbifold group is ${\bf Z}_2$. The numbers of fixed points in the twisted sectors, which are related to the number of chiral families, are always even in this case. Hence even number of families. This argument is by no means meant to be rigorous, but merely to illustrate the matter.

{}It is then only natural to consider $k=3$ models. The orbifold action in this case is ${\bf Z}_3$, and one might hope to obtain models with 3 families as the numbers of fixed points in the twisted sectors are some powers of 3. The level-3 model building appears to be more involved than that for level-2 constructions. The latter are facilitated by existence of the $E_8\otimes E_8$ heterotic superstring in 10 dimensions which explicitly possesses ${\bf Z}_2$ outer automorphism symmetry of the two $E_8$'s. Constructing a level-2 model then can be carried out in two steps: first one embeds the grand unified gauge group $G$ in each of the $E_8$'s, and then performs the outer automorphism ${\bf Z}_2$ twist. In contrast to the $k=2$ construction just sketched, $k=3$ model building requires explicitly realizing ${\bf Z}_3$ outer automorphism symmetry which is not present in 10 dimensions. The implication of the above discussion is that one needs relatively simple rules to facilitate model building. Such rules have been derived \cite{KT} within the framework of {\em asymmetric orbifolds} \cite{vafa}.

{}With the appropriate model building tools available, it became possible to construct \cite{kt,KT} and classify \cite{class} 3-family grand unified string models within the framework of asymmetric orbifolds in perturbative heterotic string theory. Here we briefly discuss the results of this classification. For each model we list here there are additional models connected to it via classically flat directions \cite{class}.

{}The following models have been found:\\
$\bullet$ One $E_6$ model with 5 left-handed and 2 right-handed families, and asymptotically free $SU(2)$ hidden sector with 1 ``flavor''.\\ 
$\bullet$ One $SO(10)$ model with 4 left-handed and 1 right-handed families, and $SU(2)\otimes SU(2)\otimes SU(2)$ hidden sector which is {\em not} asymptotically free at the string scale.\\
$\bullet$ Three $SU(6)$ models:\\
({\em i}) The first model has 6 left-handed and 3 right-handed families, and asymptotically free $SU(3)$ hidden sector with 3 ``flavors''.\\
({\em ii}) The second model has 3 left-handed and no right-handed families, and asymptotically free $SU(2)\otimes SU(2)$ hidden sector with matter content consisting of doublets of each $SU(2)$ subgroup as well as bi-fundamentals.\\
({\em iii}) The third model has 3 left-handed and no right-handed families, and asymptotically free $SU(4)$ hidden sector with 3 ``flavors''. This model has not been explicitly given in Ref \cite{class} and will be presented in Ref \cite{review}. It is a minimal $SU(6)$ extension of the minimal $SU(5)$ model \cite{Georgi}. That is, it has only 3 left-handed ${\bf 15}$'s and 6 left-handed ${\overline {\bf 6}}$'s of $SU(6)$, but no additional vector-like ${\bf 6}+{\overline {\bf 6}}$ pairs. (We note that the second $SU(6)$ model does possess extra vector-like ${\bf 6}+{\overline {\bf 6}}$ pairs, so it is {\em not} minimal in this sense even though it has only 3 left-handed but no right-handed families. Note that the number of families for $SU(6)$ is determined by the number of ${\bf 15}$'s.)\\
$\bullet$ Finally, there are some additional $SU(5)$ models which we will not discuss here in any detail for they do not seem to be phenomenologically too appealing (see below). 

{}All of the above models share some common phenomenological features. Thus, there is only one adjoint and no other higher dimensional Higgs fields in all of these models. The $E_6$ and $SO(10)$ models (and the ones connected to these via classically flat directions) do {\em not} possess anomalous $U(1)$. All three $SU(6)$ models listed above {\em do} have anomalous $U(1)$. The above models all possess non-Abelian hidden sector. There, however, exist models
(among, say, those connected to these via classically flat directions) where the hidden sector is completely broken.

{}To study phenomenological properties of these models it is first necessary to deduce tree-level superpotentials for them. This turns out to be a rather non-trivial task as it involves understanding scattering in asymmetric orbifolds. There, however, are certain simplifying circumstances here due to the fact that asymmetric orbifold models possess enhanced discrete and continuous gauge symmetries. Making use of these symmetries the tools for computing tree-level superpotentials for a class of asymmetric orbifold models (that includes the models of interest for us here) have been developed in Ref \cite{KST}. The perturbative superpotentials for 3-family grand unified string models have thus been computed in Refs \cite{KST,review}. This has made it possible to address certain phenomenological issues such as proton stability (and, more concretely, doublet-triplet splitting problem, and $R$-parity violating terms), as well as Yukawa mass matrices \cite{kstv,review}. Also, the question of supersymmetry breaking can also be addressed now \cite{review} by augmenting the tree-level superpotential with non-perturbative contributions which are under control in $N=1$ supersymmetric field theories \cite{Seiberg}.

{}Thus, in Ref \cite{kstv} doublet-triplet splitting problem and Yukawa mass matrices were studied for the $SO(10)$ models. The results of that investigation indicate that certain degree of fine-tuning would be required to solve the doublet-triplet splitting problem, suppress dangerous $R$-parity violating terms, and achieve realistic Yukawa mass matrices in the $SO(10)$ models.
Note that in the $SU(5)$ models fine-tuning seems to be unavoidable for achieving the doublet-triplet splitting as there are no ``exotic'' higher dimensional Higgs fields among the light degrees of freedom in these models, and the latter are required by all known field theory solutions to the problem. For $SU(6)$ models similar analyses have also been completed and will soon appear \cite{review}. There results of these analyses indicate that although the doublet-triplet splitting does not seem to be as big of a problem for the $SU(6)$ models as for their $SO(10)$ and $SU(5)$ counterparts, the troubles with $R$-parity violating terms and Yukawa mass matrices persist for these models as well.

{}Since none of the three-family grand unified string models constructed to date appear to be phenomenologically flawless, one naturally wonders whether there may exist other such models with improved phenomenological characteristics even within perturbative heterotic superstring vacua. There seem to exist rather compelling arguments (which are by no means proofs, however) that within free-field realized perturbative heterotic superstring models the classification of Ref \cite{class} is complete. Nonetheless, there may exist non-free-field 
grand unified string models with three families. Tools for constructing such models are way underdeveloped at present, so that for years to come the asymmetric orbifold models we have been discussing here might be the only ones available. Regardless of their phenomenological viability they provide the {\em proof of existence} for three-family grand unified string models, and could serve as stringy paradigm for such model building in general giving insight to the ``bottom-up'' approach. 

{} This work was supported in part by the grant NSF PHY-96-02074, and the DOE 1994 OJI award. We would also like to thank Albert and Ribena Yu for financial support.

\end{document}